\documentclass[aps,prl,reprint,a4paper,superscriptaddress,amsmath,amsfonts,amssymb, longbibliography]{revtex4-1}
\usepackage{bm}
\usepackage[caption=false]{subfig}
\usepackage{tikz}
\usetikzlibrary{shapes.misc,positioning,shapes}
\usetikzlibrary{patterns}
\usepackage{array, multirow}
\usepackage[colorlinks,linkcolor=blue,citecolor=blue]{hyperref}%
\usepackage{relsize}

\usepackage{times}

\begin{document}
\title{Deep Reinforcement Learning Control of Quantum Cartpoles}
\date{\today}

\author{Zhikang T. Wang}
\email{wang@cat.phys.s.u-tokyo.ac.jp}
\affiliation{Department of Physics and Institute for Physics of Intelligence, University of Tokyo, 7-3-1 Hongo, Bunkyo-ku, Tokyo 113-0033, Japan}

\author{Yuto Ashida}
\affiliation{Department of Applied Physics, University of Tokyo, 7-3-1 Hongo, Bunkyo-ku, Tokyo 113-8656, Japan}

\author{Masahito Ueda}
\affiliation{Department of Physics and Institute for Physics of Intelligence, University of Tokyo, 7-3-1 Hongo, Bunkyo-ku, Tokyo 113-0033, Japan}
\affiliation{RIKEN Center for Emergent Matter Science (CEMS), Wako, Saitama 351-0198, Japan}

\begin{abstract}
	We generalize a standard benchmark of reinforcement learning, the classical cartpole balancing problem, to the quantum regime by stabilizing a particle in an unstable potential through measurement and feedback. We use state-of-the-art deep reinforcement learning to stabilize a quantum cartpole and find that our deep learning approach performs comparably to or better than other strategies in standard control theory. Our approach also applies to measurement-feedback cooling of quantum oscillators, showing the applicability of deep learning to general continuous-space quantum control.
\end{abstract}
\maketitle

\paragraph{Introduction ---}
Over the last few decades, quantum control has attracted increasing attention owing to rapid experimental developments \cite{quantumControlSurvey,quantumControlScience}. Compared with classical control problems, there are far fewer results known for the quantum cases due to the intrinsic complexity of quantum mechanics which makes analytic approaches difficult except for simple situations \cite{spinOptimal,feedbackOptimalQuadratic,refereeCitation}. Therefore, numerical algorithms are often used instead to search for appropriate controls, such as GRAPE, QOCT and CRAB \cite{GRAPE,QuantumOptimalControlFirst,QuantumOptimalControlTheory,CRAB}. However, these methods are gradient based and only guarantee the local optimality of their strategies \cite{humanOutperformNumericalMethods,geneticQuantumControl}, and they mostly work for isolated quantum systems which are unitary and deterministic
; for stochastic systems, there is no known generally applicable approach. Thus, it is important to explore alternative versatile strategies, and this is where machine learning is expected to be effective. 

Deep reinforcement learning (RL) is a cutting-edge machine learning strategy that uses deep learning in its RL system. It is model-free, requires no prior knowledge, and often achieves state-of-the-art performance. Recently, deep RL has been applied to a few quantum control problems, including the manipulation of spin systems \cite{spinChainControl}, finding robust and precise control of qubits \cite{universalQubitControlDenoise,semiconductorDotDeepRLControl,alphaZeroSearchControl} and designing quantum error correcting gates \cite{ errorCorrectionGate,errorCorrectionOptimization}. In most of these cases, deep RL achieves success by demonstrating performance comparable or superior to conventional methods \cite{spinChainControl,universalQubitControlDenoise,alphaZeroSearchControl}. In addition, it deals with problems that are intractable by existing methods \cite{errorCorrectionGate,errorCorrectionOptimization,semiconductorDotDeepRLControl}. Especially, RL has been applied to the control problems that are not amenable to analytical approaches \cite{spinChainControl,multipleGateControl,errorCorrectionGate,semiconductorDotDeepRLControl}. 

Despite its success, the full potential of deep RL is yet to be explored; in particular, all of the previous studies have only focused on discrete systems. Indeed, there are \textit{continuous-space} cases that need control, including superconducting circuits and cavity optomechanical systems, and various transport problems such as in trapped ion systems \cite{superconductingQubitAnharmonicityPulseControl,optomechanics,shortcutsToAdiabaticity}. Continuous-space systems are typically harder to control, since they have infinitely many levels and control Hamiltonians usually do not commute with the system Hamiltonian. Therefore, it is unclear whether deep RL can handle the continuous case, and if so, to what extent compared with existing methods. In this Letter, as a proof of principle, we demonstrate that deep RL can indeed solve simple continuous-space control problems, even in the presence of measurement backaction noise. Although Ref.~\cite{QLearningKapitzaOscillator} also considers a continuous-space system, it focuses on a finite control sequence and uses the primitive Q-table learning strategy which has limited applicability \cite{RLSutton}. To the best of our knowledge, the present work is the first attempt to apply deep RL to continuous-space quantum control.
\begin{figure}[b]
	\subfloat{
		\begin{tikzpicture}
		\pattern[thin, pattern=north east lines] (-45pt,-3pt) rectangle (45pt,0);
		\draw[thick] (-45pt,0pt) -- (45pt,0pt);
		\draw[thick] (-18pt,10pt) rectangle (18pt, 26pt);
		\draw[thick] (-9.5pt,5pt) circle [radius=5pt];
		\draw[thick] (9.5pt,5pt) circle [radius=5pt];   
		\draw[thick, latex-latex] (-15pt,-7pt) --node[below]{control} (15pt,-7pt);
		\draw[thin,fill=black,rotate around={-10:(0pt,19pt)}] (-0.4pt,19pt) rectangle (0.4pt, 60pt) node[thick,draw,fill=lightgray,circle,inner sep=3pt]{};
		\draw[thick,dashed](0pt,19pt)-- ++(0pt,45pt);
		\node(label) at (-40pt,52pt) {(a)};
		\end{tikzpicture}
	}
	\subfloat{	\begin{tikzpicture}
		\node[inner sep=0pt] (quantum cartpole) at (0,0)
		{\centering
			\subfloat{
				\includegraphics[width=0.4\linewidth]{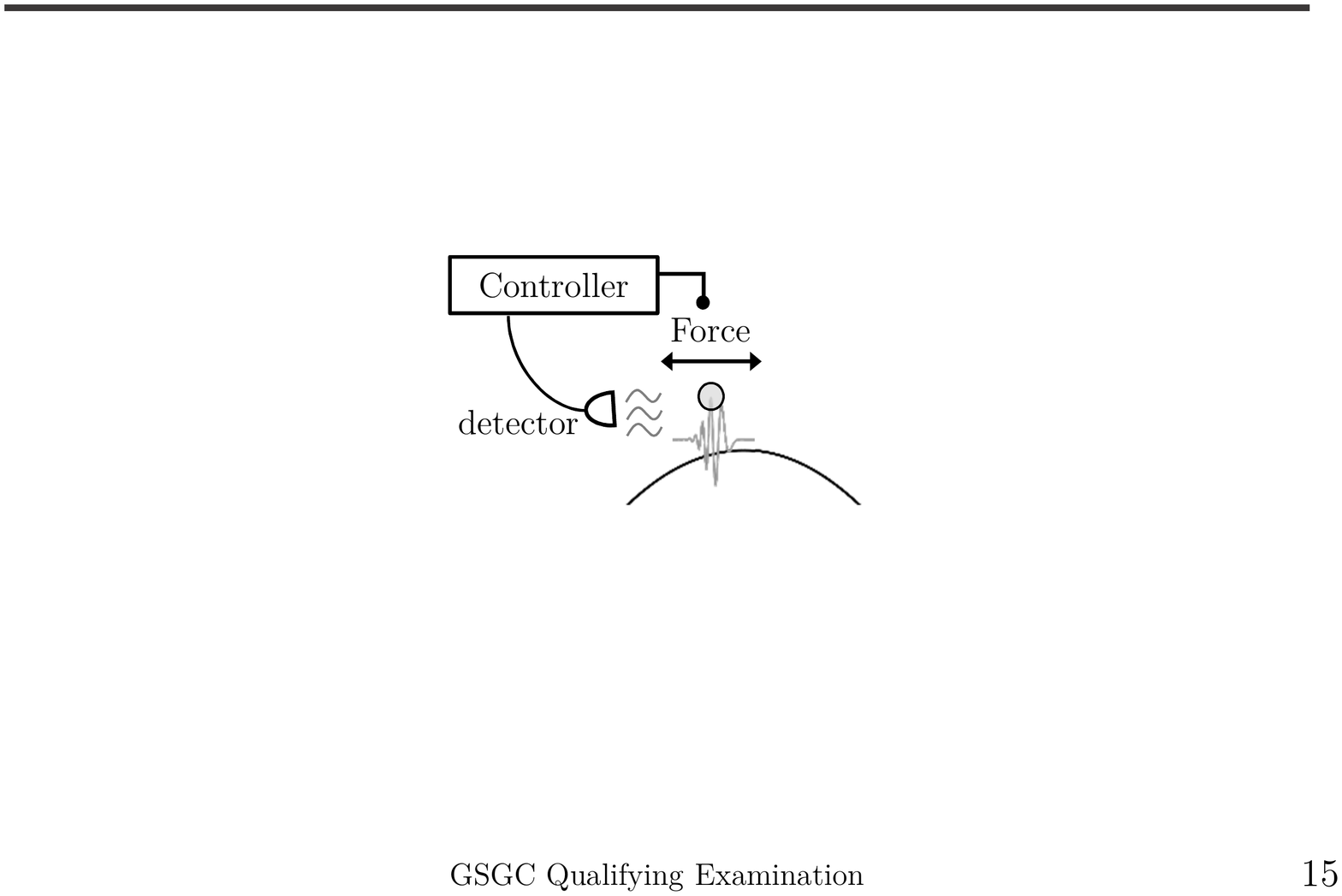}}};
		\node(label) at (-55pt,34.5pt) {(b)};
		\end{tikzpicture}
	}
	\caption{\label{cartpoles}(a) Classical cartpole system. A controller moves the cart to prevent the pole, which is an inverted pendulum, from falling down. (b) Generalized quantum cartpole system. A controller monitors a particle and applies forces to keep it close to the top of a potential.}
\end{figure}

We construct quantum analogues to the classical cartpole balancing problem which is arguably the best known RL benchmark \cite{cartpole,RLbenchmarks,openAIGym}. We apply deep RL to solve the \textit{quantum cartpole} problem. In the classical cartpole problem, a controller moves a cart properly to prevent the pole from falling down [see Fig.~\ref{cartpoles}(a)], which is a prototypical example of controlled stabilization of an unstable system. In analogy, we consider measurement-feedback control of an unstable quantum system [see Fig.~\ref{cartpoles}(b)]. The quantum cartpole is more difficult to stabilize than the classical one, since quantum measurement that is needed to localize the state of a particle exerts backaction on the particle. Adopting state-of-the-art deep Q-learning \cite{DQNNature,rainbowDQN}, we find that deep RL can stabilize the quantum cartpole just like the classical one, and when quantum effects are significant, deep RL outperforms other known control strategies. We also apply our approach to the problems of measurement-feedback cooling of oscillators. The results show a great potential of deep RL for continuous-space and stochastic quantum control. From the perspective of machine learning, the quantum control problems themselves can serve as a new type of hard benchmark task for reinforcement learning AI \footnote{Our codes are released on \url{https://github.com/Z-T-WANG/DeepReinforcementLearningControlOfQuantumCartpoles}}.
\paragraph{Deep Q-Learning ---}
We use deep Q-learning as our RL strategy and implement the deep Q-network (DQN) algorithm. We briefly introduce RL and DQN in the following.

Given reward $r$, RL aims to find a policy $\pi$ that maximizes the total reward $\sum_{t}r_t$ at discretized times $t$. The reward could be the obtained score in a game, artificially assigned $\pm1$ to represent a success or failure, or some control target in a control problem. In general, RL artificial intelligence (AI) interacts with a controllable system, and the reward $r(s)$ is a function of the system state $s$ \footnote{The system controlled by the AI is termed \textit{environment} in machine learning literature. However, since this terminology conflicts with the meaning of \textit{environment} in quantum mechanics, we do not use it.}. RL starts with no prior knowledge and learns from trial and error to maximize $\sum_{t}r(s_t)$, by taking actions $a_t=\pi(s_t)$ that can influence the time evolution $s_t \to s_{t+1}$ of the system, where $\pi$ is the learned action policy.

Q-learning realizes RL through a \textit{Q function} that represents the expected total future reward of a policy $\pi$, defined by \cite{RLSutton}
\begin{equation}\label{Q def}
Q^\pi(s_t,a_t)=r(s_t)+\mathbb{E}_{\{(s_{t+i},a_{t+i})|\pi\}_{i=1}^{\infty}}\!\left[\sum_{i=1}^{\infty}\gamma^i_{q}\, r(s_{t+i})\right],
\end{equation}
where $\gamma_{q}$ is a discount factor by which the future reward will be discounted and satisfies $0\!<\!\gamma_{q}\!<\!1$, and the expectation is taken over trajectories of the action and the state of the controlled system. The parameter $\gamma_{q}$ is manually set to ensure the convergence and close to 1. For an optimal policy $\pi^*$ which maximizes $Q^{\pi^*}\!\!$, $Q^{\pi^*}$ satisfies the following Bellman equation \cite{Bellman,DQNNature}
\begin{equation}\label{Q optimal}
Q^{\pi^*}(s_t,a_t)=r(s_t)+\gamma_{q}\,\mathbb{E}_{s_{t+1}}\left[\max_{a_{t+1}}Q^{\pi^*}(s_{t+1},a_{t+1})\right].
\end{equation}
In addition, the function $Q^{\pi^*}$ that satisfies the equation is unique under general assumptions. Therefore, if we find a function that satisfies Eq.~(\ref{Q optimal}), we effectively obtain $Q^{\pi^*}$ and can thus derive the optimal policy $\pi^*$. This is the basic idea of Q-learning.

Deep Q-learning uses a deep feedforward neural network as a universal function approximator to approximate $Q^{\pi^*}$ \cite{deeplearningNature,deepLearningBook}, and the network is called a deep Q network (DQN) and denoted by $f_{\bm{\theta}}(s,a)$, where $\bm{\theta}$ represents the internal parameters and the system state $s$ is the network input. Assuming that the space of actions $a$ is discrete, the DQN outputs a value $f_{\bm{\theta}}$ for each choice of $a$ at its output layer. The function $f_{\bm{\theta}}(s,a)$ approximates $Q^{\pi^*}$ by optimizing its parameters $\bm{\theta}$ so that the left-hand side of Eq.~(\ref{Q optimal}) becomes approximately equal to the right-hand side. In this way, $f_{\bm{\theta}}(s_t,a_t)$ approximately solves Eq.~(\ref{Q optimal}), and it is used to represent $Q^{\pi^*}$ and to find the optimal policy $\pi^*$. As suggested in Ref.~\cite{rainbowDQN}, we incorporate the state-of-the-art technical advances of deep Q-learning, which include prioritized sampling, noisy networks, double Q-learning and the duel network structure \cite{prioritizedSampling,NoisyDQN,DoubleDQN,DuelDQN}. See Ref.~\cite{rainbowDQN} for details.
\paragraph{Quantum cartpole ---}
As shown in Fig.~\ref{cartpoles}(a), the classical cartpole is a simple system that is stabilized by an external control. Instead of quantizing this two-body cartpole, we consider a one-body one-dimensional system that reproduces its stability properties. Specifically, we put a particle at the top of a potential and try to keep it at that unstable position using appropriate external forces. The Hamiltonian is
\begin{equation}\label{Hamiltonian}
\hat{H}(F)=\frac{\hat{p}^2}{2m}+V(\hat{x})-F\hat{x},
\end{equation}
where $V$ is the potential and $F$ is a controllable time-dependent force. We require $V$ to be symmetric about $0$ and $V\to-\infty$ for $x\to\pm\infty$, and we require $|F|$ to be bounded from above by $F_{\text{max}}$. It is clear that under unitary evolution, the wave function cannot be stabilized at the top of the potential due to delocalization. We perform continuous position measurement on the particle \cite{continuousMeasurement}, so that the wave function contracts due to state reduction and can be stabilized by means of measurement-based feedback control. The density operator $\rho$ of the system is governed by the following It\^o stochastic differential equation: 
\begin{equation}\label{mixed time-evolution equation}
d\rho=-\frac{i}{\hbar}[\hat{H},\rho]dt  -\frac{\gamma}{4}[\hat{x},[\hat{x},\rho]]dt
+\sqrt{\frac{\gamma\eta}{2}}\{\hat{x}-\langle \hat{x}\rangle,\rho\}dW,
\end{equation}
where $dW$ is a Wiener increment sampled from the Gaussian distribution $\mathcal{N}(0,dt)$, $\gamma$ is the measurement strength, $\eta\in[0,1]$ is the measurement efficiency, and $\{\cdots,\cdots\}$ is the anticommutator.

In the classical cartpole problem, if the tilting angle of the pole exceeds a certain threshold, we judge that the pole has fallen down. Similarly, we judge that the stabilization of the quantum cartpole fails if more than 50\% of the probability distribution of the particle lies outside certain boundaries $\pm x_{\text{th}}$. Then, a controller attempts to keep the cartpole from this failure for as long time as possible. Because of stochastic measurement backaction, an open-loop control generally fails and a measurement-feedback control is necessary. The controller may either take the raw measurement outcomes or the postmeasurement state as its source of information. For deep RL, this becomes the input $s$ of the neural network. 

To implement Q-learning, we discretize the interval $\left [-F_{\text{max}}, F_{\text{max}}\right ]$ of the control force into 21 equispaced choices, which correspond to the actions $a$, and we discretize the time into control steps. At each control step, the controller decides a force $F$, and the force is kept constant until the next step. Concerning the RL reward, we follow the original cartpole problem and choose $r=1$ if the system is stable and $Q=r=0$ if it fails, so that the RL aims at infinitely long-time stabilization.

As a proof of principle, we investigate two simple cases of $V$, i.e., an inverted harmonic potential $V=-\frac{k}{2}\hat{x}^2$ and an inverted quartic potential $V=-\lambda \hat{x}^4$. It is known that a particle in a quadratic potential under continuous position measurement is described by a Gaussian Wigner distribution \cite{CoherentStatesByDecoherence,LinearStochasticSchrodingerEquation}. This is because the shape of its Wigner distribution is preserved by the time evolution, and its Gaussianity monotonically increases with the measurement. Thus the state is fully characterized by the means $\langle\hat{x}\rangle,\langle\hat{p}\rangle$ and the covariances $C_{xx},C_{pp},C_{xp}$, where $C_{xx}:=\langle\hat{x}^2\rangle-\langle\hat{x}\rangle^2$, $C_{pp}:=\langle\hat{p}^2\rangle-\langle\hat{p}\rangle^2$ and $C_{xp}:=\frac{1}{2}\langle\hat{x}\hat{p}+\hat{p}\hat{x}\rangle-\langle\hat{x}\rangle\langle\hat{p}\rangle$. As the covariances converge to steady-state values during the time evolution \cite{feedbackOptimalQuadratic}, the state becomes effectively described by $\langle\hat{x}\rangle$ and $\langle\hat{p}\rangle$, and the effective time-evolution equations are
\begin{subequations}\label{quadratic time-evolution}
	\begin{eqnarray}
	d\langle\hat{x}\rangle&=&\frac{\langle \hat{p}\rangle}{m}dt+\sqrt{2\gamma\eta}\,C_{xx}\,dW,\\
	d\langle\hat{p}\rangle&=&(-k{\langle \hat{x}\rangle}+F)dt+\sqrt{2\gamma\eta}\,C_{xp}\,dW,
	\end{eqnarray}
\end{subequations}
where $dW$ is the Wiener increment in Eq.~(\ref{mixed time-evolution equation}). In contrast, in a quartic potential, a Gaussian state is not preserved due to nonlinearity; moreover, we have $d\langle\hat{p}\rangle=-4\lambda\langle\hat{x}^3\rangle dt\neq-4\lambda\langle\hat{x}\rangle^3 dt$, while a classical particle should obey $dp=-4\lambda x^3dt$. Thus, the system is fully characterized by the expectation values and exhibits intrinsic quantum-mechanical behavior. In fact, it is known that the quartic system corresponds to the $\phi^4$ theory in quantum field theory \cite{anharmonicphi4}, and therefore it is interesting to investigate how the quartic system can be controlled by RL.

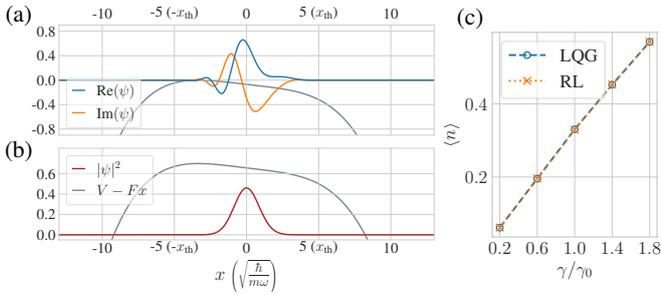
\begin{figure}[t]
	\begin{tikzpicture}
	\node[inner sep=0pt] (quantum cartpole) at (-10pt,0pt)
	{\centering
		\includegraphics[width=0.63\linewidth]{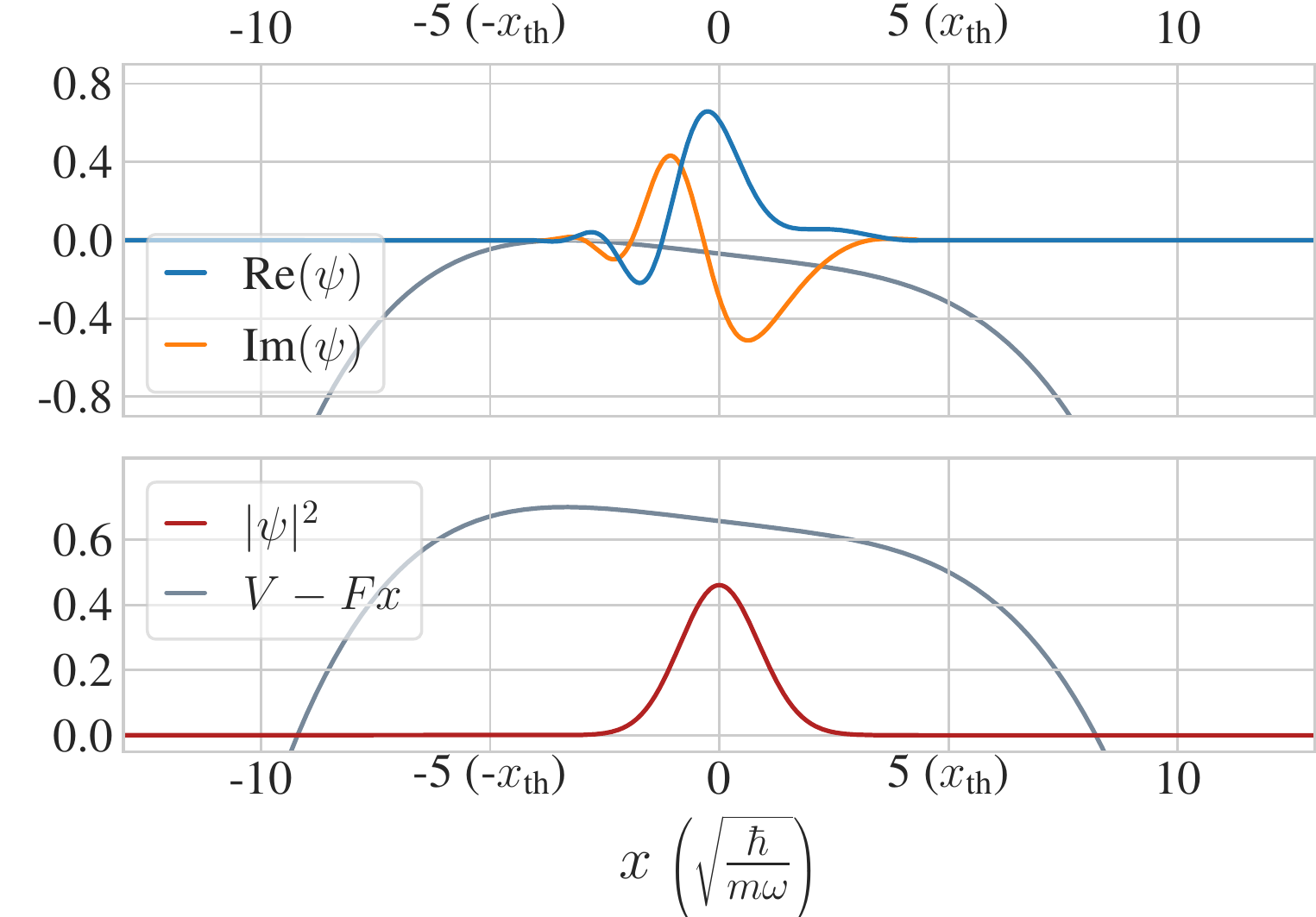}};
	\node(a) at (-88pt,50pt) {(a)};
	\node(b) at (-88pt,-1pt) {(b)};
	\node[inner sep=0pt] (gamma plot) at (112pt,-2pt)
	{\centering
		\includegraphics[width=0.33\linewidth]{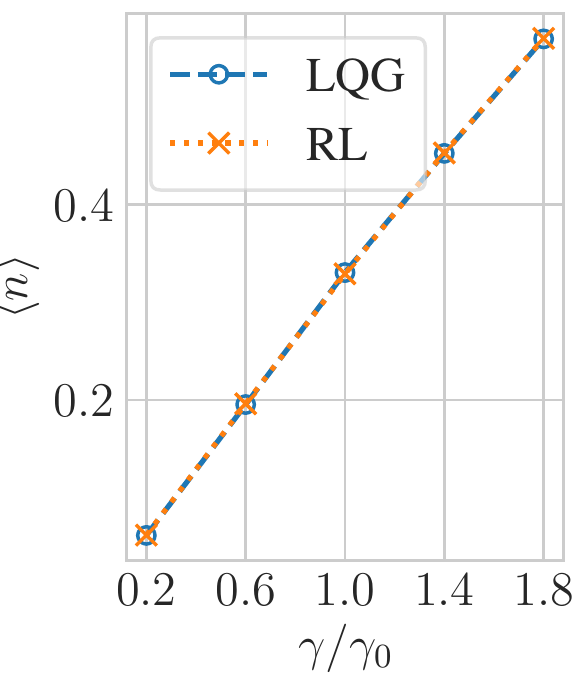}};
	\node(c) at (81pt,49pt) {(c)};
	\end{tikzpicture}	\vspace{-7mm}
	\caption{\label{control plot}(a),(b) Snapshots of the controlled quartic cartpole system. The blue and the orange curves are the real and imaginary parts of the wave function, and the grey and red curves are the controlled potential $V-Fx$ and the probability density. The scale of the potential is arbitrary. See Supplemental videos \cite{Note1}. (c) Performance of the deep RL and the LQG control on the task of cooling harmonic oscillators, plotted against varying normalized measurement strength $\gamma/\gamma_0$, where $\gamma_0$ is a default value.}
\end{figure}
Now we determine the system parameters to numerically simulate the systems and train the RL controller. For simplicity, we assume the perfect measurement efficiency $\eta=1$ and we only consider pure states. With measurement outcomes, an initial mixed state is purified and we can compute and obtain the trajectory-level quantum state, and therefore the wave function is available even when we consider an experimental setup. The measurement strength $\gamma$ is set such that the wave function has a width comparable to the ground state of the noninverted potential. The control strength $F_{\text{max}}$ is determined so that the peak of the controlled potential $V-Fx$ is allowed to move within several times the width of the wave function, and the failure threshold $x_{\text{th}}$ is set to be the peak position of $V+F_{\text{max}}x$, as shown in Fig.~\ref{control plot}. The number of control steps is roughly 30 per oscillation period of the noninverted potential. These settings ensure that the cartpole problems are nontrivial and there are nonvanishing probabilities of stabilization failure due to measurement backaction. Concerning the input $s$ of the neural network $f_{\bm{\theta}}$, we try three different choices: (i) the sequence of measurement outcomes in time; (ii) the wave function $|\psi\rangle$; (iii) the distribution moments such as $\langle\hat{x}\rangle,\langle\hat{p}\rangle$, $C_{xx}$, $\langle\left(\hat{x}-\langle\hat{x}\rangle\right)^3\rangle$. The length of the measurement outcome sequence and the high-order cutoff for the distribution moments are determined so that the AI learns easily. The setting of RL hyperparameters mainly follows that given in Ref.~\cite{rainbowDQN}. Specific parameter values and details are given in the Supplemental Material.

To benchmark the above RL quantum control scheme, we first test it on the measurement-feedback cooling problems with the corresponding noninverted potentials ($+\frac{k}{2}x^2$ and $+\lambda x^4$). The RL reward becomes the minus system energy, and all other settings remain the same.
\paragraph{Results ---}
We train the RL AI on the measurement-feedback cooling problems of quantum harmonic and quartic oscillators and on the stabilization problems of quantum harmonic and quartic cartpoles. For each problem, about $5\times10^5$ oscillation periods are simulated to carry out the training. The resulting dynamics of the controlled quantum systems are plotted and recorded as videos in the Supplemental Material \cite{Note1}, clearly showing that the RL control is nontrivial. To evaluate its performance, we benchmark and compare it with other control strategies. All the control strategies we consider are discretized in the same manner to allow for fair comparison. 
\begin{table}[t]
	\begin{tabular}{p{4.2em}p{6em}cc}\hline
		Controller &Input& Cooling ($\langle\hat{n}\rangle$) & Cartpole control ($T$)\\ \cline{1-4}
		\multirow{5}{3.7em}{Deep RL} & Distribution moments& $\mathbf{0.329\pm0.001}$ & $\mathbf{41.71\pm0.32}$ \\ \cline{2-4}
		& Wave function& $0.330\pm0.001$ & $38.44\pm0.30$\\ \cline{2-4}
		& Measurement outcomes& $0.372\pm0.001$ & $12.77\pm0.20$ \\ \cline{1-4}
		\multicolumn{2}{c}{LQG control} & $\mathbf{0.331\pm0.001}$ & $\mathbf{42.02\pm0.33}$\\ 
		\cline{1-4}
		\multicolumn{2}{c}{No control} & Heats up to $\infty$ & $0.52\pm0.01$ \\ \hline
	\end{tabular}
	\caption{\label{harmonic results}Results of deep RL control for harmonic systems as a benchmark, i.e.~cooling of the oscillator and stabilization of the cartpole, compared with the LQG control which is optimal for the harmonic case. The performances are measured in terms of the average number of excitations $\langle\hat{n}\rangle$ and the average time before failure in units of the oscillation period $T$, with the estimation error.}
\end{table}

For the measurement-feedback cooling problem of harmonic oscillators, using the Gaussian approximation, the optimal control is given by the standard linear-quadratic-Gaussian (LQG) control theory \cite{feedbackOptimalQuadratic,LQG}. Therefore, we benchmark our deep RL control with this optimal one, and compare the average energy of their controlled systems. As the LQG controller minimizes the squared position and momentum of a particle, it can also be used to stabilize a cartpole, and therefore we also compare our results with that obtained by the LQG control. The results are listed in Table~\ref{harmonic results}. Since the performance of the deep RL using the measurement outcomes as input is not good, we do not consider this method for the quartic problems.

For quartic oscillators, the optimal control is unknown and thus we compare the RL controller with simple control strategies, including controlled damping, the LQG control, and a control designed by us which makes use of semiclassical approximation. Note that the greedy strategy is included as a special case of controlled damping as it always minimizes the kinetic energy of the particle. The damping and the LQG coefficients are determined by grid search to obtain the best results. The control based on semiclassical approximation assumes the Gaussianity and a fixed variance $C_{xx}$ of the wave function so as to map $\langle\hat{x}\rangle$ and $\langle\hat{p}\rangle$ to the position $x$ and the momentum $p$ of a classical particle, and it uses the optimal control for the derived classical particle. The obtained results are presented in Table~\ref{anharmonic results}.

As shown in Table~\ref{harmonic results} and \ref{anharmonic results}, the deep RL successfully solves the control problems. It can match the performance of the optimal control, and when no optimal control is known, it can outperform other strategies, demonstrating itself as a strong candidate for quantum control. To confirm that the results also hold for other parameter settings, we take the harmonic oscillator as an example and vary its measurement strength $\gamma$ and repeat the numerical experiments, using distribution moments as the input for the AI. The results are plotted in Fig.~\ref{control plot}(c). Also to confirm its robustness, we test the performance of the trained AI for imperfect measurement efficiency, and we indeed find that the performance only gradually degrades with a decrease in the measurement efficiency. See the Supplemental Material for details.

We find that the LQG control also performs well for most of the cases. This is because the state is well localized and Gaussian-like due to the harmonic potential or the strong measurement in the cartpole systems, which allows for effective classical control. For the cooling problem of quartic oscillators, in the presence of significant non-Gaussianity, none of the conventional controllers performs well. In contrast, the model-free deep RL deals with general potentials and handles both Gaussian and non-Gaussian states well, thus showing its versatility and superiority over the other approaches.
\paragraph{Discussion ---}
Although deep RL can achieve prominent results, its performance depends on inputs of the neural network. This is presumably because the network cannot handle its input information precisely. For measurement outcome inputs, the neural network does not distinguish between recent and distant past measurement outcomes and learn all of them equally, which may hinder learning. For the wave function inputs, since physical observables are of the form $\langle\psi|\hat{O}|\psi\rangle$ that is quadratic in the wave function $|\psi\rangle$, the neural network using linear mappings and rectified linear units may not accurately evaluate the relevant physical quantities. Therefore, it seems that there is room for improvement on the AI side, and within the current framework of deep RL, well-tuned AI can be helpful to achieve the best results on the problems. 
\begin{table}[t]
	\begin{tabular}{p{4.2em}p{6em}cc}\hline
		Controller &Input& Cooling ($\langle \hat{H}\rangle-E_0$) & Cartpole ($T$)\\ \cline{1-4}
		\multirow{3}{3.7em}{Deep RL} & Distribution moments& $\mathbf{0.0057\pm0.0001}$ & $\mathbf{13.84\pm0.21}$ \\ \cline{2-4}
		& Wave function& $0.0065\pm0.0001$ & $12.89\pm0.20 $ \\ \cline{1-4}
		\multicolumn{2}{c}{Controlled damping} & $0.0169\pm0.0005$ & $2.32\pm0.03$ \\[2pt]
		\multicolumn{2}{c}{LQG control} & $0.0140 \pm 0.0005$ & $\mathbf{13.45\pm0.21}$ \\[2pt]
		\multicolumn{2}{c}{Semiclassical control} & $\mathbf{0.0113\pm0.0003}$ & $5.32\pm0.08$ \\ \cline{1-4}
		\multicolumn{2}{c}{No control} & Heats up to $\infty$ & $0.81\pm0.01$\\ \hline
	\end{tabular}
	\caption{\label{anharmonic results}Results of deep RL control for quartic systems, compared with other strategies. The performance is evaluated in terms of the average energy $\langle \hat{H}\rangle$ subtracted by the ground-state energy $E_0$. The quartic systems are constructed to be comparable to the harmonic ones and the units are the same as those in Table~\ref{harmonic results}.}
\end{table}

Despite its versatility, deep RL also has several limitations. Especially, as our AI learns from numerically simulated quantum trajectories to control the real systems, to apply our approach, we must have a reliable and simulatable physical model for the controlled experimental system in order to train the AI, and the computational cost for the simulation and training must be affordable. Further details on computational cost are given in the Supplemental Material. The measurement and feedback considered in this Letter also require real-time evaluation of the trajectory-level states and fast computation of the neural network to realize the control in experiments.

An important future extension of this work is to consider mixed states. If the numerical experiments on mixed states are successful, the AI controller can be applied to wider experimental systems such as cavity optomechanical systems as in Ref.~\cite{measurementFeedbackCavityOptomechanics}. Among the most important directions is to apply the deep RL approach to control complicated and realistic systems, which may improve the performance of current quantum devices. From the perspective of RL, the quantum control problems can serve as RL benchmarks, since the problems are qualitatively different from most of the existing RL tasks. The quantum tasks are difficult, stochastic, and yet of practical significance, while most of the current RL benchmarks are obtained only for deterministic toy models or simple video games \cite{openAIGym, Atari}.
\paragraph{Conclusion ---}
We have constructed quantum analogues to the classical cartpole balancing problem using inverted harmonic and quartic potentials. We have used deep RL to tackle the quantum cartpole problems and measurement-feedback cooling of the corresponding quantum oscillators. The systems are numerically simulated and are stochastic and continuous space. We have demonstrated that the deep RL can match or outperform other strategies in these problems, showing a great potential of deep RL for general continuous-space quantum control, and these quantum control tasks may also serve as RL benchmarks \cite{Note1}.
\begin{acknowledgments}
	The authors thank Ryusuke Hamazaki for fruitful discussions. This work was supported by a Grant-in-Aid for Scientific Research (KAKENHI Grant No.~JP18H01145) and a Grant-in-Aid for Scientific Research on Innovative Areas Topological Materials Science (KAKENHI Grant No.~JP15H05855) from the Japan Society for the Promotion of Science. Z.~T.~W. is supported by Global Science Graduate Course (GSGC) program of the University of Tokyo. Y.~A. acknowledges support from the Japan Society for the Promotion of Science through Grant No. JP16J03613.
\end{acknowledgments}

\bibliography{references}
\widetext
\pagebreak
\begin{center}
	\textbf{\large Supplemental Material}
\end{center}

\subsection{Parameters Used in Numerical Experiments}
The parameters used in our numerical experiments are listed in Tables~\ref{harmonic} and \ref{quartic}. Units thereof are shown in the parentheses, where a reference mass $m_c$ and a reference angular frequency $\omega_c$ are used. The potentials of the harmonic system and the quartic system are respectively given by $V=\frac{k}{2}x^2$ and $V=\lambda x^4$. The period $T$ of the harmonic oscillator is $\frac{2}{\omega_c}$.
\begin{table}[h]
	\begin{tabular}{p{4.3em}ccccc}\hline
		& $\omega$ ($\omega_c$) & $m$ ($m_c$) & $k$ ($m_c\omega^2_c$) & $\gamma$ ($\frac{m_c\omega_c^2}{\hbar}$) & $F_{\text{max}}$ ($\sqrt{\hbar m_c\omega_c^{3}}$)\\ \cline{1-6}
		harmonic oscillator & $\pi$ & $\dfrac{1}{\pi}$ & $\pi$ & $\pi$ & $5\pi$ \\ 
		quadratic cartpole & N/A & $\dfrac{1}{\pi}$ & $-\pi$ & $2\pi$ & $8\pi$ \\ \hline
	\end{tabular}
	\caption{\label{harmonic}Parameters of the quadratic systems used in the numerical experiments.}
\end{table}
\begin{table}[h]
	\begin{tabular}{p{4.3em}cccc}\hline
		& $m$ ($m_c$) & $\lambda$ $\left (\frac{m^2_c\omega^3_c}{\hbar}\right )$ & $\gamma$ ($\frac{m_c\omega_c^2}{\hbar}$) & $F_{\text{max}}$ ($\sqrt{\hbar m_c\omega_c^{3}}$) \\ \cline{1-5}
		quartic oscillator & $\dfrac{1}{\pi}$ & $\dfrac{\pi}{25}$ & $\dfrac{\pi}{100}$ & $5\pi$ \\ 
		quartic cartpole & $\dfrac{1}{\pi}$ & $-\dfrac{\pi}{100}$ & $\pi$ & $5\pi$ \\ \hline
	\end{tabular}
	\caption{\label{quartic}Parameters of the quartic systems used in the numerical experiments.}
\end{table}\\
The control boundaries $x_{\text{th}}$ of the quadratic cartpole and the quartic cartpole systems are set at $8$ and $5 \left (\sqrt{\frac{\hbar}{m_c\omega_c}}\right )$, respectively, and the measurement efficiency $\eta$ is assumed to be unity. The above setting is the default in our released codes \cite{Note1}.
\subsection{Details of Implemented Controllers}
The control force is piecewise constant and it is changed once per $\frac{T}{36}$, which we call a control step. The range $[-F_{\text{max}},F_{\text{max}}]$ is discretized into 21 equispaced values from which the control force of each controller is chosen.

The linear-quadratic-Gaussian (LQG) controller takes two input values $x$ and $p$ at the beginning of a control step and chooses a force in an attempt to satisfy the condition $p=-\sqrt{m|k|}\,x$ at the end of the control step. This controller effectively minimizes the quadratic control loss $\int(\frac{p^2}{2m}+\frac{kx^2}{2})\,dt$ and is known to be optimal in the presence of Gaussian noise in the continuous limit \cite{LQG}. Here we have removed the control loss associated with the force $F$ to make a fair comparison with our reinforcement learning (RL) controller. For quantum systems, the expectation values $\langle\hat{x}\rangle$ and $\langle\hat{p}\rangle$ are used in place of the classical position $x$ and momentum $p$.

The damping controller takes $p$ as an input value and attempts to reduce $p$ exponentially by changing $p$ to $(1-\zeta)p$ at each control step.

The semiclassical controller assumes the Gaussianity of the state, which implies the zero skewness $\langle(\hat{x}-\langle\hat{x}\rangle)^3\rangle=0$ and the zero excess kurtosis $\frac{\langle(\hat{x}-\langle\hat{x}\rangle)^4\rangle}{C^2_{xx}}-3=0$, where $C_{xx}:=\langle\hat{x}^2\rangle-\langle\hat{x}\rangle^2$. Therefore, for the quartic systems, we have 
$$\lambda\langle\hat{x}^4\rangle=\lambda(6C_{xx}\langle\hat{x}\rangle^2+\langle\hat{x}\rangle^4+3C^2_{xx})$$
Assuming a fixed $C_{xx}$, we replace $\langle\hat{x}\rangle$ and $\langle\hat{p}\rangle$ with classical position $x$ and momentum $p$ and set $V=6C_{xx}\lambda x^2+\lambda x^4$. Then we seek to minimize $\int(\frac{p^2}{2m}+|V|)\,dt$, which is the average energy of a quartic oscillator. If the system is deterministic and free from noise, this minimization is achieved if the condition $p=-\sqrt{|2m(6\lambda C_{xx}+\lambda x^2)|}x$ is satisfied. Thus, the semiclassical controller tends to make the controlled system satisfy this above condition at each control step.

When the LQG controller is applied to quadratic systems, $k$ and $m$ are simply the quadratic system parameters. The same holds true for the semiclassical controller on quartic systems. However, when we use the LQG controller and the damping controller on quartic systems, we perform a small grid search to find the optimal parameters $\zeta$ and $k$ for the controllers to perform well and we use the best parameter choices that we found. The parameter values are included in our released codes \cite{Note1}.
\subsection{Robustness Test}
As the deep RL controller is trained only with perfect measurement information and noiseless input, we need to confirm that the controller is robust against noise and imperfection. We consider the case where the trained controller is used with imperfect measurement, i.e.~efficiency $\eta<1$. We keep the measurement strength $\gamma$ unchanged, but only a fraction $\eta$ of the original measurement outcomes are made available to the controller. The controller then ``mistakenly" computes the wave function of the controlled particle assuming that all the measurement outcomes are already collected by it and that the measurement strength is $\eta\gamma$. This setup represents a situation in which the environment can monitor the particle but we ignore such effects by simply assuming that the total measurement strength is $\gamma$. Then, the ``mistakenly" computed wave function is used to obtain the input for the AI controller. We emphasize that the AI has only been trained on the perfect measurement setting. Therefore, this setup can add persistent noise to the AI input.

We test the deep RL controllers that input distribution moments, on all the four tasks mentioned in the paper. The results are plotted below in Fig.~\ref{eta plot}. The fact that the performances gradually decrease with decreasing $\eta$ shows that the deep RL controller is indeed robust. The performance is dramatically different from the case of no control (see main text).
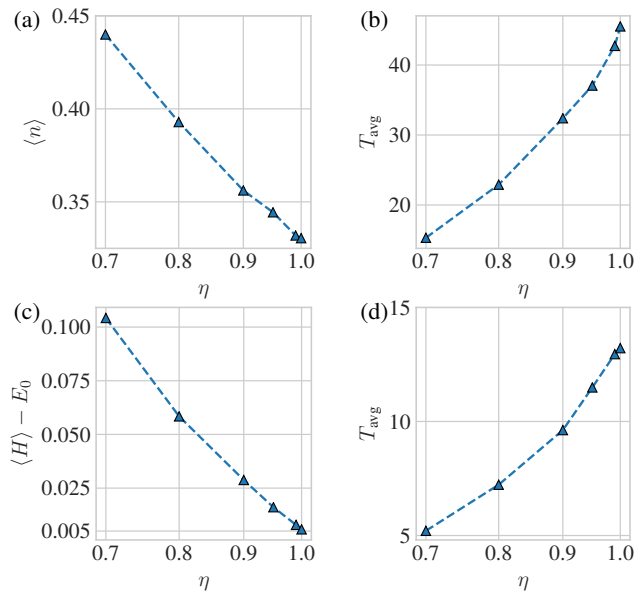
\begin{figure}[ht]
	\begin{tikzpicture}
	\node[inner sep=0pt] (harmonic cartpole) at (45pt,0pt)
	{\centering
		\includegraphics[height=0.215\linewidth]{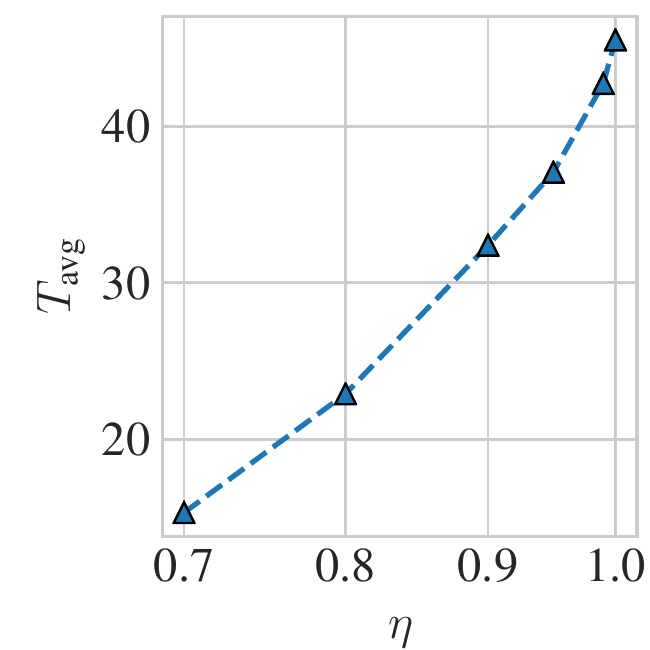}};
	\node(b) at (2pt,50pt) {(b)};
	
	\node[inner sep=0pt] (harmonic cooling) at (-75pt,0pt)
	{\centering
		\includegraphics[height=0.215\linewidth]{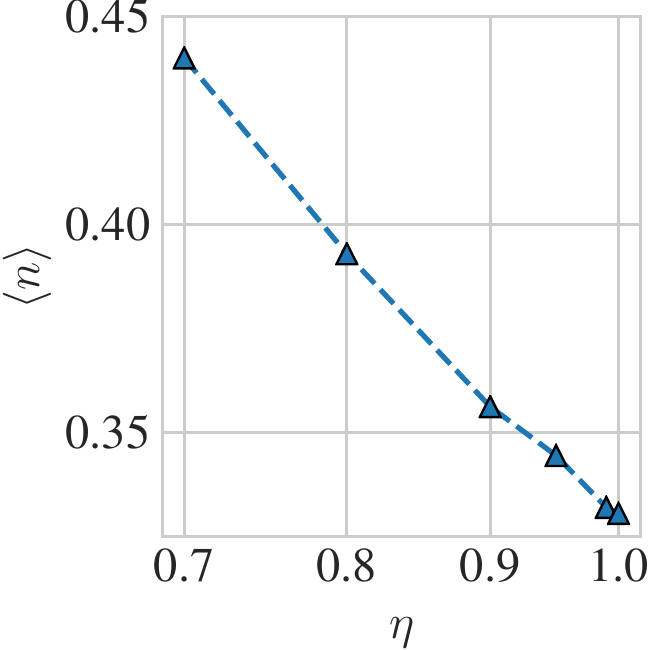}};
	\node(a) at (-128pt,50pt) {(a)};
	\node[inner sep=0pt] (quartic cartpole) at (45pt,-110pt)
	{\centering
		\includegraphics[height=0.215\linewidth]{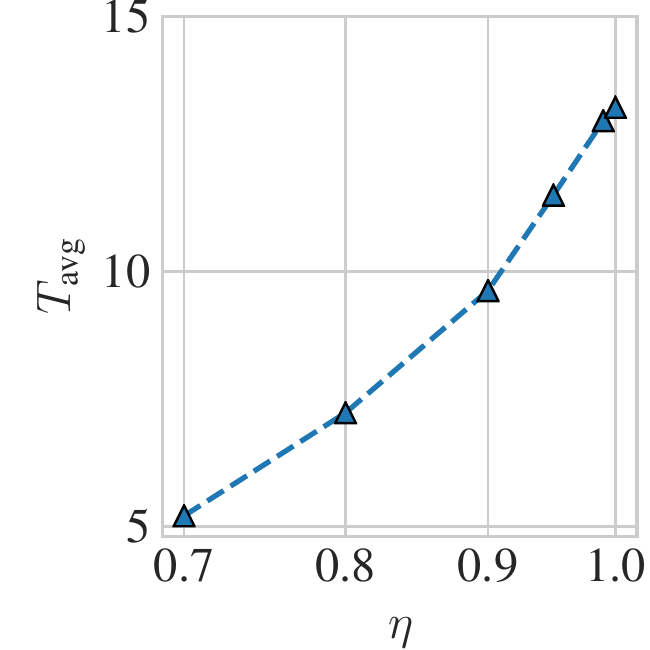}};
	\node(d) at (2pt,-60pt) {(d)};
	
	\node[inner sep=0pt] (quartic cooling) at (-78pt,-110pt)
	{\centering
		\includegraphics[height=0.215\linewidth]{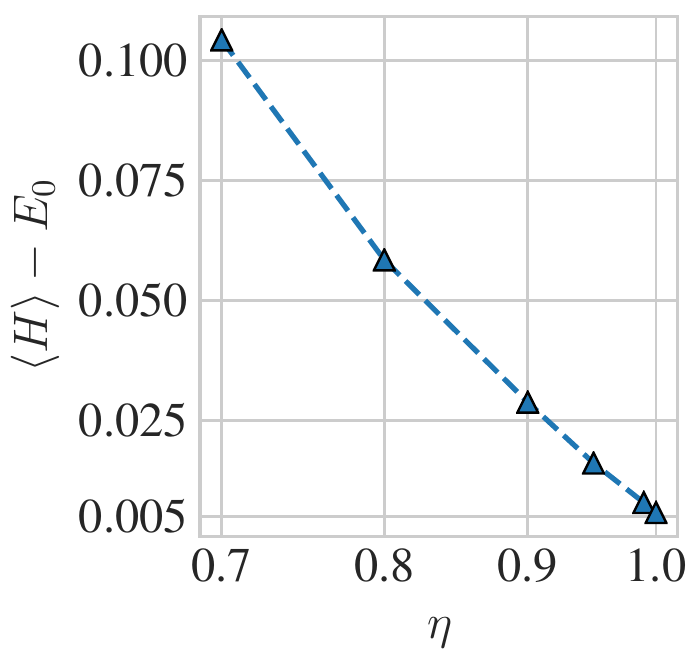}};
	\node(c) at (-128pt,-60pt) {(c)};
	\end{tikzpicture}
	\caption{\label{eta plot}Performance with imperfect measurement efficiency $\eta$. (a) and (b) show the harmonic cooling and harmonic cartpole tasks corresponding to Table I in the main text, and (c) and (d) show the quartic cooling and quartic cartpole tasks corresponding to Table II in the main text.}
\end{figure}
\subsection{Computational Cost}
The computational cost of the deep RL approach is mostly determined by the cost of the quantum simulation involved, as the training and the evaluation of the neural networks are typically much less costly. In RL, the controller do trials and error to understand the effect of a control on the controlled system, and therefore to accumulate enough experience to learn, it needs to try many times and consequently, the controlled quantum system also needs to be simulated for a long time. In our numerical experiments, we simulate the quantum systems for about $5\times10^5 T$ to complete the training, and this is the major computational cost in our numerical experiments. As $T$ is approximately the oscillation period of the relevant systems, for other quantum systems, we also expect that the simulation of about $5\times10^5$ times the typical time scale of the controlled system would be needed for the AI controller to be trained. Therefore, for quantum systems that are numerically expensive to simulate, the deep RL approach may not be a viable choice. Nevertheless, because the numerical bottleneck lies in the part of accumulating empirical data, the algorithm can be highly parallelized. We expect that it can be parallelized even up to hundreds of, or probably thousands of processes, which can alleviate the numerical bottleneck and speed up the algorithm substantially. In our implementation codes \cite{Note1}, we have used around 10 to 40 parallelized processes, which speeds up the algorithm for tens of times.

\begin{figure}[ht]
	\begin{tikzpicture}
	\node[inner sep=0pt] (quartic cartpole) at (-20pt,-120pt)
	{\centering
		\includegraphics[height=0.239\linewidth]{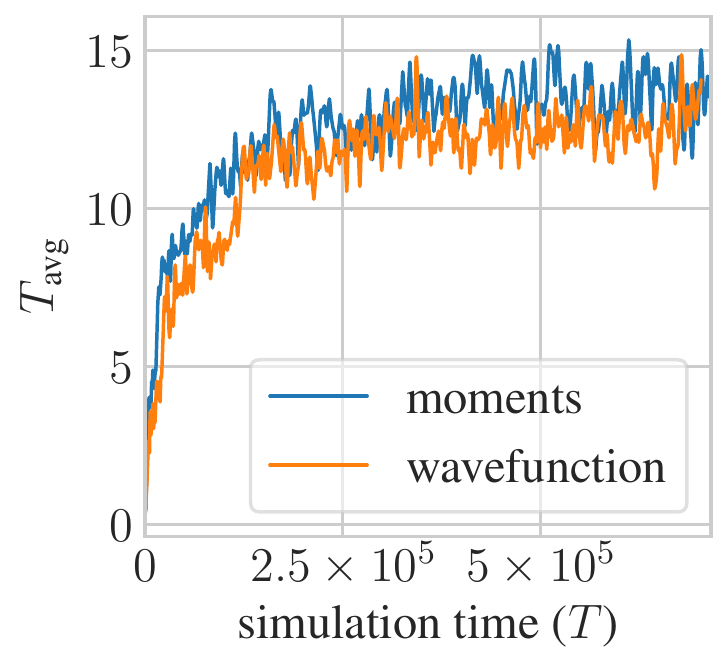}};
	\node(c) at (-80pt,-70pt) {(c)};
	
	\node[inner sep=0pt] (quartic cooling) at (49pt,0pt)
	{\centering
		\includegraphics[height=0.215\linewidth]{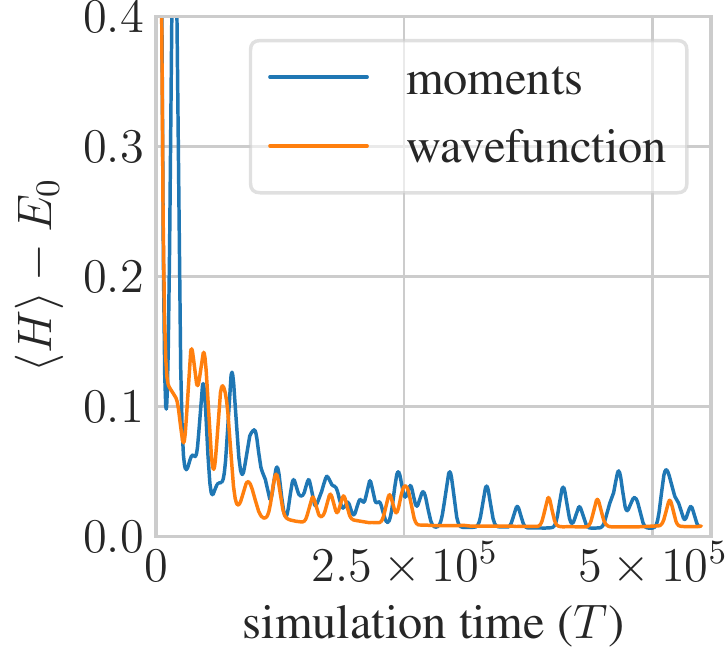}};
	\node(b) at (-5pt,47pt) {(b)};
	
	\node[inner sep=0pt] (harmonic cooling) at (-75pt,0pt)
	{\centering
		\includegraphics[height=0.215\linewidth]{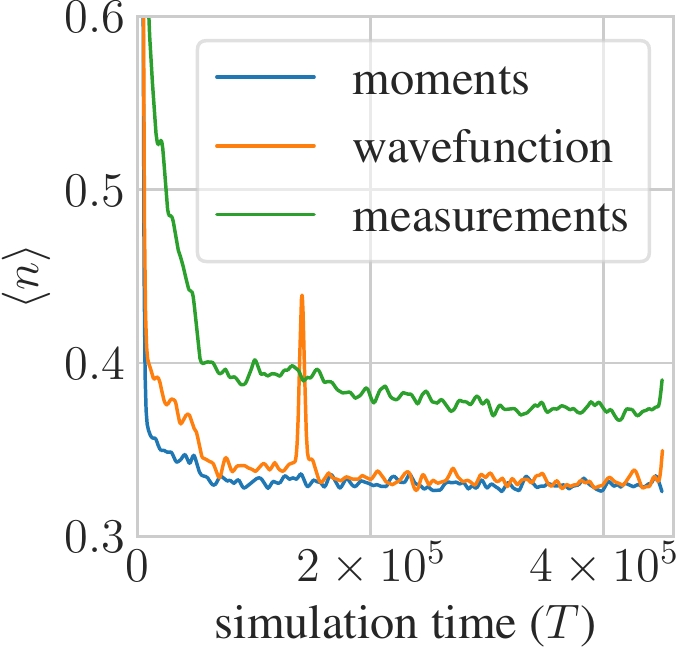}};
	\node(a) at (-128pt,47pt) {(a)};

	\end{tikzpicture}
	\caption{\label{learning curves}Learning curves of the RL controller for (a) the harmonic cooling task, (b) the quartic cooling task, and (c) the quartic cartpole task. We see that performance changes steeply at the early stage (at or before $10^4 T$); until then, the control is trivial, the system energy diverges, and the stabilization time $T_{\text{avg}}$ diminishes. Smoothing is applied to the curves.}
\end{figure}
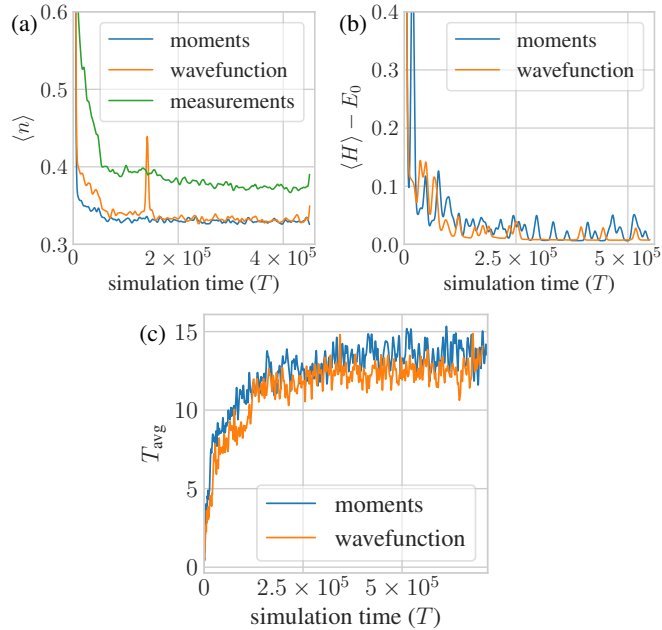

Meanwhile, one does not need the full $5\times10^5 T$ simulation to make the AI start to learn. The AI starts learning and improves its performance largely at an early stage of training, and in our experiments it only marginally improves after an early stage. This suggests that we can qualitatively know whether the deep RL approach works or not without the need of complete training. As an example, we present the learning curves of our training in the following as in Fig.~\ref{learning curves}. We see that the AI has qualitatively changed performances at the early stage, when the simulation time is only around $10^4$.
\subsection{Settings of Reinforcement Learning}
All the settings mentioned below can be found and confirmed in our released implementation codes \cite{Note1}. The deep Q networks used in this research are the standard feedforward neural networks with linear connections and rectified linear unit (ReLU) activations. The neural networks that input distribution moments or wave function data are composed of 4 fully connected layers, where the last two layers are separated into two branches following Ref.~\cite{DuelDQN}. The numbers of neurons in the 4 layers are respectively 512, 256, 256+128 and 21+1 for the tasks of a harmonic potential, and for the tasks of a quartic potential the 2nd layer is changed to contain 512 neurons. Wider and deeper neural networks have higher capability in general, but they are also harder to start learning in an RL setting. The networks that input measurement outcomes are composed of 3 one-dimensional convolutional layers and 3 fully connected layers. The convolution kernel sizes and strides are (13,5), (11,4) and (9,4), and the numbers of filters are 32, 64 and 64. The neurons in the fully connected layers are 256, 256+128 and 21+1. The last two layers of the networks are always noisy layers using factorized noise (see Ref.~\cite{NoisyDQN}), and other fully connected layers learn normalized weight matrices as suggested in Ref.~\cite{weightNormalization}.

For the cooling tasks, the reinforcement learning reward is the scaled minus energies of the controlled systems. For the cartpole tasks, the reward is always $10$ when the system does not fail, and when the system fails, the expected future reward $Q$ is set to be zero. To ensure that the neural networks have output values of a moderate size, we rescale the received reward by a factor of $1-\gamma_{q}$ when we train the network.

When training on the cooling tasks, we discard the accumulated experiences that are associated with high system energies; otherwise the network may not learn. This is because a high-energy experience typically results in a large training loss, which disturbs the learning of an appropriate control for low energy cases, especially at an early stage of training. The energy cutoff is $\langle\hat{n}\rangle=10\sim20$ for the harmonic oscillator and $12\hbar\omega_c$ for the quartic oscillator.

The networks input measurement outcome sequences of time duration $\frac{3}{2}T$ for the harmonic oscillator problem and $2T$ for the quadratic cartpole problem, using 2880 measurement outcome values per duration $T$. Note that the previously applied control forces should also be a part of the input information together with the measurement outcomes; otherwise the controlled state is not fully determined. The networks that input wave function data simply separate the complex wave function into real and imaginary parts and use them as the network inputs. The input distribution moments include up to the second moment for the harmonic system and up to the fifth moment for the quartic system.

The reinforcement learning actors adopt the $\epsilon$-greedy strategy to take action and gather experience for training \cite{DQNNature}. The training algorithm uses the double Q learning strategy as in Ref.~\cite{DoubleDQN}, and the update period of target networks is set to be 300 times the gradient descent step. Specifically, we first set the update period to a small number at the start of training, and when the learning proceeds, we gradually increase it to 300. The gradient descent algorithm we use is a modified version of Adam which we call Laprop \cite{Adam, Ziyin}. The size of the memory replay buffer is about $3\times10^5T$ for the cases of distribution moments and wave function inputs, and for the case of measurement outcome inputs, the size is about $6\times10^4T$. When the memory replay buffer becomes full, we discard previous experiences randomly. The batch size is 512, and each experience is learned 8 times on average. We use Pytorch as our deep learning library and use its default initialization for network parameters \cite{Pytorch}. The Q-learning parameter $\gamma_q$ is 0.99. We also apply a learning rate schedule and rescaling of the input where appropriate. See the released codes for details \cite{Note1}. We find that fine-tunings of the RL strategy and the parameters often cause a change in performance, suggesting that the quantum control problems are useful RL benchmarks that provide meaningful and insightful results. 
\subsection{Evaluation of Performance}
The quantum systems are initialized as small Gaussian wave packets with zero momentum at the center, except for the quartic oscillator system. For the quartic oscillator, we implement a two-stage initialization by first initializing a Gaussian state with a small momentum and then letting it evolve for a random time duration between $\frac{15}{2}T\sim10T$. In this manner, we obtain a sufficiently non-Gaussian state, and we use this state as the initial state of the quartic oscillator that is to be controlled.

After the systems are initialized, we apply different controls and record the system behavior. For a cartpole system, we simply record how much time elapses before the system fails, and we repeat this procedure sufficiently many times to obtain an estimate for the average control time to use as the performance. We find that the recorded results approximately follow an exponential distribution, and the variance of the estimates is large. For the cooling problems, we record the energies of the controlled systems as a measure of the performance. To alleviate the effect of initialization, for the harmonic oscillator, we start to record the system energy at $15T$ after initialization, and we start to record the quartic system energy at $25T$ after initialization. The systems are simulated up to $50T$ and then reinitialized.

To produce the results in the main text, we first collect the trained AI models during the training process, and then we do a validation test to pick out the best-performing model among the collected models. Finally we do the test again on the best-performing model and report the performance in the final test, which is the standard validation-and-test procedure in machine learning.

The ground-state energy $E_0$ of the quartic oscillator shown in Table~II of the main text is approximately $0.2285\hbar\omega$, and the first and the second excited state energies are $0.8186\hbar\omega$ and $1.6063\hbar\omega$, which are obtained by exact diagonalization of the numerically simulated system. Because the measurement squeezes the wave function in position space, it is impossible for the state to continue to stay at the ground-state energy, and the lowest possible energy under control is always larger than $E_0$. However, we do not know a lower bound.
\end{document}